\documentclass[11pt]{article}
\usepackage{moriond,epsfig}
\def\be{\begin{equation}}
\def\ee{\end{equation}}
\def\bea{\begin{eqnarray}}
\def\eea{\end{eqnarray}}

\begin{document}
\vspace*{4cm}
\title{PHYSICS AT THE INTERNATIONAL LINEAR COLLIDER}

\author{ J. LIST }

\address{DESY, Notkestr. 85,\\
22607 Hamburg, Germany}

\maketitle\abstracts{The International Linear Collider (ILC) is the next large project in accelerator based particle physics. It is complementary to the LHC in many aspects. Measurements from both machines together will finally shed light onto the known deficiencies of the Standard Model of particle physics and allow to unveil a possible underlying more fundamental theory. Here, the possibilities of the ILC will be discussed with special emphasis on the Higgs sector and on topics with a strong connection to cosmological questions like extra dimensions or dark matter candidates.
}

\section{Introduction}
The Standard Model of particle physics (SM) provides a unified and precise description of all known subatomic phenomena. It is consistent at the quantum loop level and it covers distances down to $10^{-18}$~m and times from today until $10^{-10}$~s after the Big Bang. Despite its success, the SM has some deficiencies which indicate that it is only the effective low energy limit of a more fundamental theory. These deficiencies comprise the absence of experimental evidence for the Higgs particle, its number of free parameters and their values, fine--tuning and stability problems above energies of about 1~TeV, and, last but not least, its ignorance of gravity. Furthermore, the SM contains no particle which could account for the cold dark matter observed in the universe.

There are good reasons to expect phenomena beyond the SM at the TeV scale, i.e. in the reach of the immediate generation of new accelerators. Any new physics which solves the hierarchy problem between the electroweak and the Planck scale needs to be close to the former. Experimental hints arise from the fits to electroweak precision data, which require either a Higgs boson mass below 250~GeV or something else which causes similar loop corrections. Furthermore, most cold dark matter scenarios based on the hypothesis of a weakly interacting massive particle favour masses of about 100~GeV.

If there are new particles ``around the corner", the LHC is likely to discover them. The ultimate goal, however, is not only to discover new particles, but to measure their properties and interaction with high precision in order to pin down the underlying theory and to determine its parameters. In the unlikely case that the LHC will not find any new particles, the task of the ILC would be to measure the SM parameters with even higher precision than before in order to find out what is wrong with todays fits that point to a light Higgs and new physics at the TeV scale. In any case, an electron position collider will provide an invaluable tool complementary to the LHC. 

\section{The Accelerator}
Due to the synchrotron radiation, whose power increases with the energy to the forth power, storage rings become impractical for electrons or positrons at energies significantly higher than those achieved at LEP. Thus, two linear accelerators, whose beams collide in the center, are the only way to realise electron positron collisions at center of mass energies of 500~GeV - 1~TeV. There is a worldwide consensus to build such a linear collider with the following parameters: The center of mass energy should be tunable between 200 and 500~GeV and upgradable to up to 1~TeV. The integrated luminosity for the first 4 years should be 500~fb$^{-1}$ and the beams should be polarised to 80\% for the electrons and up to 60\% for the positrons. There should be two interaction regions and the experiments should run concurrently with the LHC.

In the 1990ies, several linear collider projects were developped; the first one to publish a technical design report including a full costing was the TESLA project in 2001. In the following years there was a competition between designs based on warm or superconducting technology.
In 2004, an international agreement was achieved to use the superconducting technology, because of its high energy efficiency, the less stringent alignment tolerances, its use for the planned X-FEL at DESY and because its functionality has been demonstrated at the TESLA Test Facility at DESY. Since 2005, the Global Design Effort, lead by Barry Barish from Caltech, is coordinating the further design of the ILC. Its first achievement was a new common definition of the baseline parameters of the accelerator. The next steps will be the costing of the current design by the end of 2006 and a full engineering until 2008.

For a high luminosity machine as the ILC, also the experiments have to fulfill high demands on precision. Thus, in parallel to the design of the accelerator, the development of the different detector components and the optimisation of their combination is pursued with high intensity.

\section{Higgs Physics}
The last unobserved ingredient of the Standard Model of particle physics is the Higgs boson. It gives mass to the originally massless weak gauge bosons and to all quarks and charged leptons, and it is essential to keep the theory finite. The most characteristic property of this fundamental scalar field is that its couplings have to be proportional to the mass of the particles it interacts with -- if the interaction with the Higgs field is indeed responsible for the mass generation.

If the SM Higgs boson exists, it will be discovered at the LHC once the data from about one year of running time have been analysed.
The LHC experiments will also be able to measure quite precisely its mass and the ratios of couplings. 

The ILC on the other hand would be a factory for Higgs bosons, with as many Higgs events expected per day as the LHC experiments collect per year. The high production rate and the clean environment of an electron positron collider will allow to determine very precisely the complete profile of the Higgs boson and thus to establish the Higgs mechanism experimentally and to prove that this new particle {\itshape really} is the SM Higgs boson. If it turns out that it is not, it might be one of several Higgs bosons appearing in alternative models, e.g. in Supersymmetry. Then it is even more important to analyse the profile of all accesible Higgs bosons in order to determine the type and the parameters of the underlying theory.

\begin{figure}[htb]
\begin{center}
\begin{picture}(8,4)
\put(0,0){\epsfig{figure=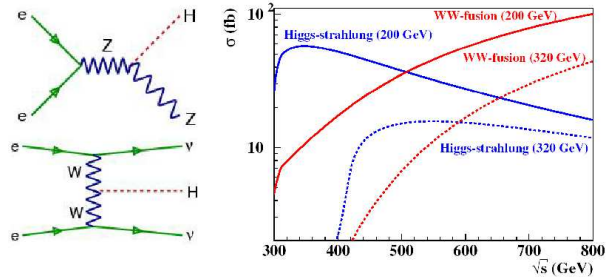,width=8cm}}
\end{picture}
\end{center}
  \caption{\label{fig:higgsfeyn} Left: Feynman diagrams for Higgsstrahlung (top) and $WW$ fusion (bottom) Right: cross sections of the two processes as function of the center of mass energy for Higgs masses of 200 and 320~GeV.}
\end{figure}

\begin{figure}[htb]
\begin{center}
\begin{picture}(20,5)
\put(1,0){\epsfig{figure=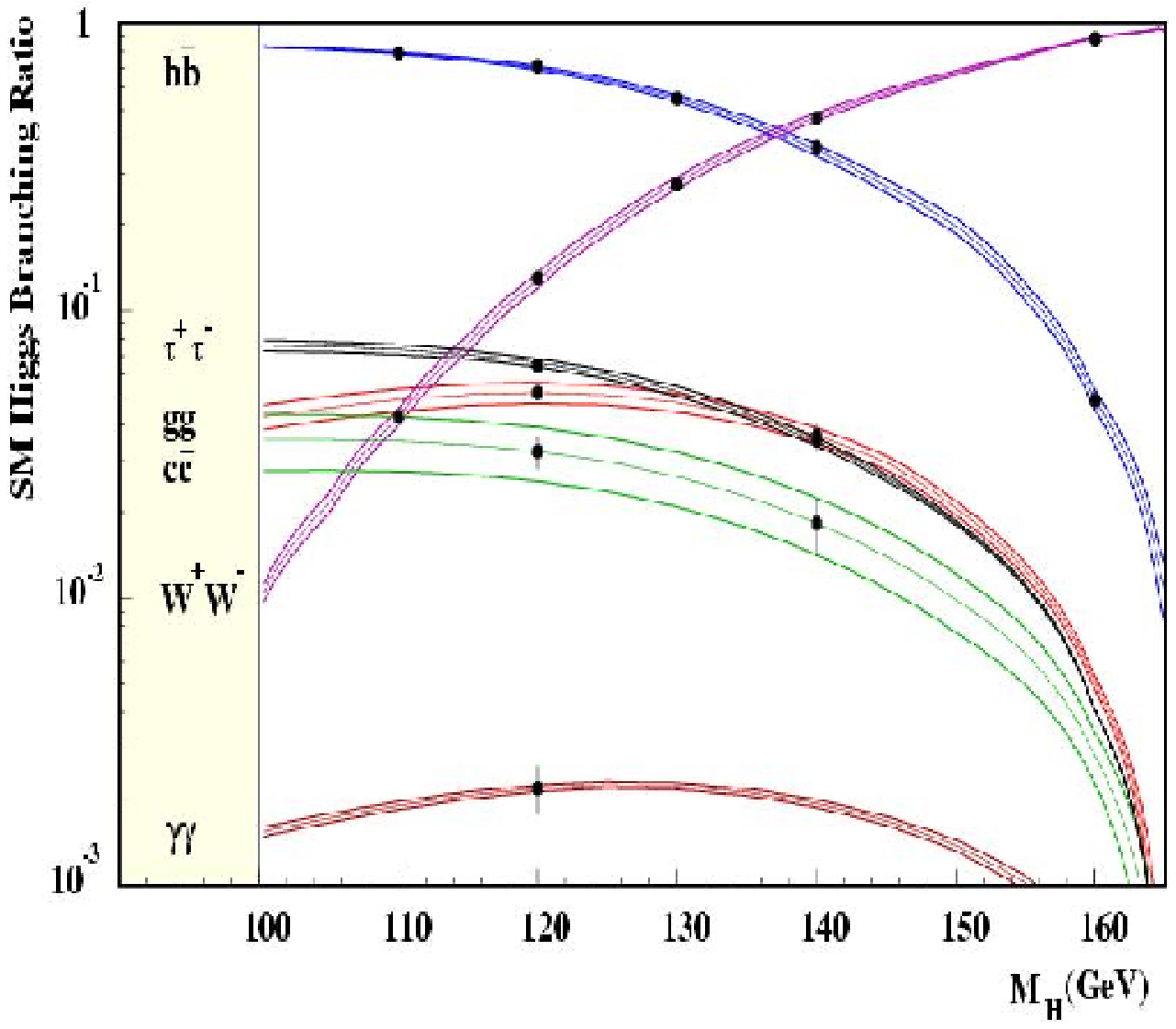,width=5.5cm}}
\put(9,0){\epsfig{figure=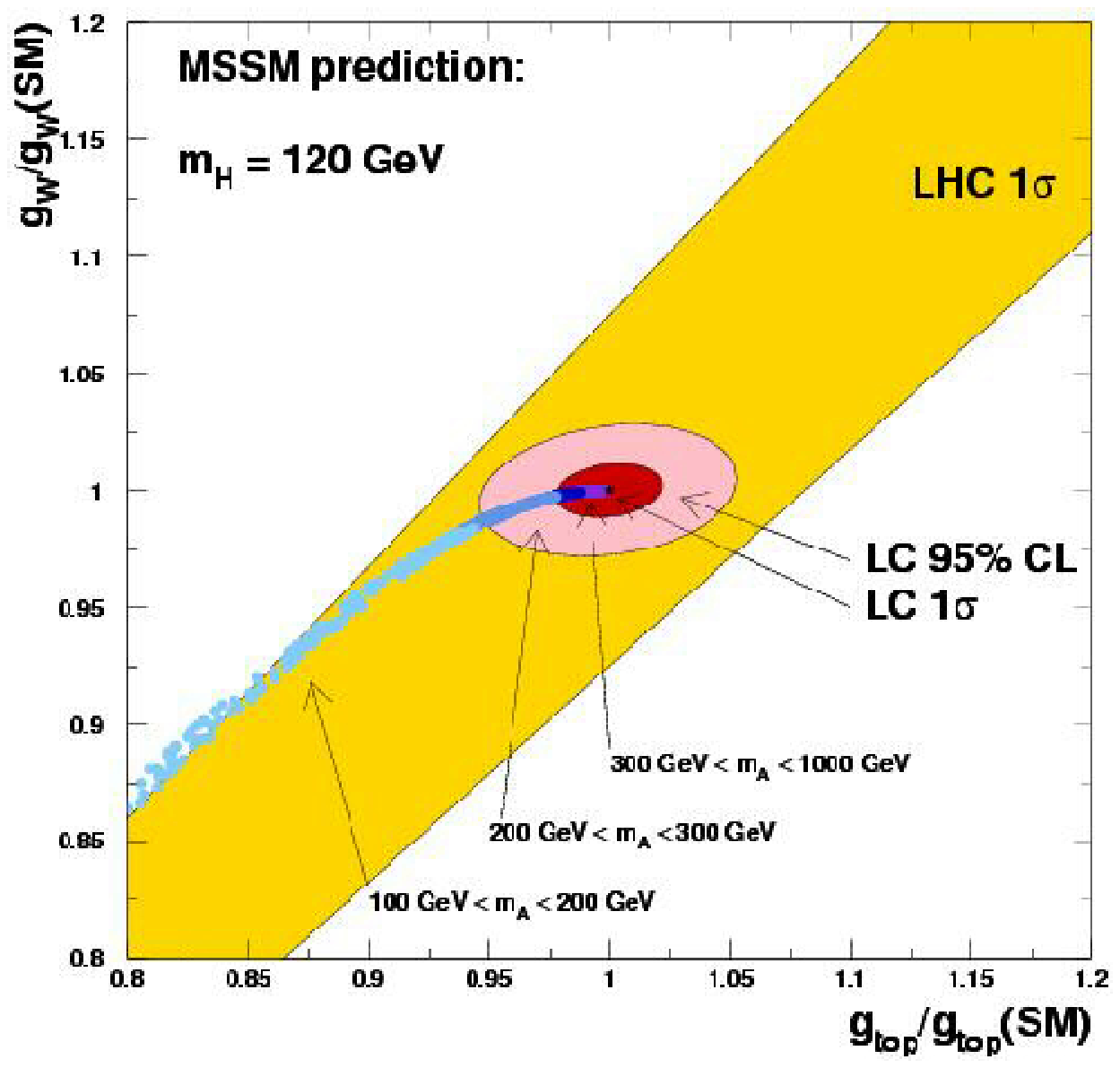,width=5cm}}
\end{picture}
\end{center}
  \caption[]{\label{fig:higgs} Left: Expected precision of the Higgs branching ratio measurements at the ILC as a function of the Higgs mass~\cite{tdr}. Right: SM and MSSM predictions for $g_{ttH}$ vs $g_{WWH}$ and expected precision of the corresponding LHC and ILC measurements~\cite{tdr}.}
\end{figure}

%
At the ILC, the SM Higgs boson will be produced by two processes Higgsstrahlung and $WW$ fusion, whose Feynman diagrams and cross sections are shown in figure~\ref{fig:higgsfeyn}. The Higgsstrahlungs process, which dominates for center of mass energies up to about 300~GeV above the Higgs mass, provides the possibility of a decay mode independent mass measurement via the recoil mass against the $Z^0$ boson produced together with the Higgs, which even works for decays to invisible particles. In addition, the mass measurement has been studied in many decay channels of the Higgs boson. With the detectors planned for the ILC, precisions below 0.1\% are achievable over the mass range favoured by the SM. The $WW$ fusion process is a unique tool for determining the total width of the Higgs boson even for low higgs masses, i.e. $m_H < 200$~GeV, by measuring the total $WW$ fusion cross section and the branching ration BR($H\rightarrow WW^*$). At high masses, the Higgs is so broad that its width can be determined directly from its lineshape. The total width then gives access to all couplings via measuring the branching fractions. Figure~\ref{fig:higgs} shows the expected precision for the different Higgs branching ratios. Especially disentangling decays to $b\bar{b}$, $c\bar{c}$ and $gg$ is challenging and requires an excelent vertex detector.

The only coupling not accessible in decays is the top Higgs Yukawa coupling $g_{t}$ due to the high mass of the top quark. At the ILC alone, one would need to collect 1000~fb$^{-1}$ at 800~GeV to produce enough $e^+e^- \rightarrow t \bar{t} H$ events to reach a precision of 5\% to 10\% for $g_{t}$. A more elegant way to extract $g_{ttH}$ would be to combine the rate measement of $q\bar{q}/gg \rightarrow t \bar{t} H$ (with the Higgs decaying further to $b\bar{b}$ or $W^+W^-$) from the LHC, which is proportional to $g_{t}^2 g_{b/W}^2$ with the absolute measurements of $g_{b}$ and $g_{W}$ from the ILC. The mass of the top quark is one of the most important SM parameters in order to check the overall consistency of the Higgs mechanism. With a scan of the production threshold, the top mass can be measured at the ILC to 50-100~MeV and its width to 3-5\%.

The ultimate proof of the Higgs mechanism is the measurement of the Higgs self coupling $\lambda$, which allows to map the Higgs potential and check the relation between $\lambda$, the Higgs mass and its vacuum expectation value: $\lambda = m_H^2/2v^2$. This needs Higgsstrahlungs events, where the Higgs itself radiates off a second Higgs boson, which leads to 6-fermion final states, including 6-jet final states. These require an excellent jet energy resolution, much superior to for example the LEP detectors. 

Once all Higgs parameters are measured, a global fit to all Higgs properties will answer the question if it really is the SM Higgs boson - or for example a supersymmetric one - even if no other new particle should be observed at the LHC. Even if other particles will have been already discovered, a careful analysis of the Higgs sector is essential to unveil the model beneath the new phenomena and to determine its parameters.
Figure~\ref{fig:higgs} shows the SM and some MSSM expectations in the $g_{t}$-$g_{W}$ plane and the precisions achievable a the LHC and the ILC.

If Supersymmetry (SUSY) is realised in nature, there will be at least five physical Higgs bosons: two CP-even bosons $h$ and $H$, similar to the SM one, a neutral, CP-odd boson $A$, and two charged Higgs bosons $H^{\pm}$. In contrast to the SM case, the Higgs masses are not free parameters, but depend in on other SUSY and SM parameters. The lightest CP-even Higgs $h$, for example, must have a mass of less than about 135~GeV. Despite its large center of mass energy, there are regions in the SUSY parameter space where the LHC can see only one of the Higgs bosons, depending on the parameters of the Higgs sector, for example in maximal mixing scenario at intermediate values of  $\tan{\beta}$. In these cases, the additional precision information from the ILC will be especially important. 

\begin{figure}[htb]
\begin{center}
\begin{picture}(12,5)
\put(0,0){\epsfig{figure=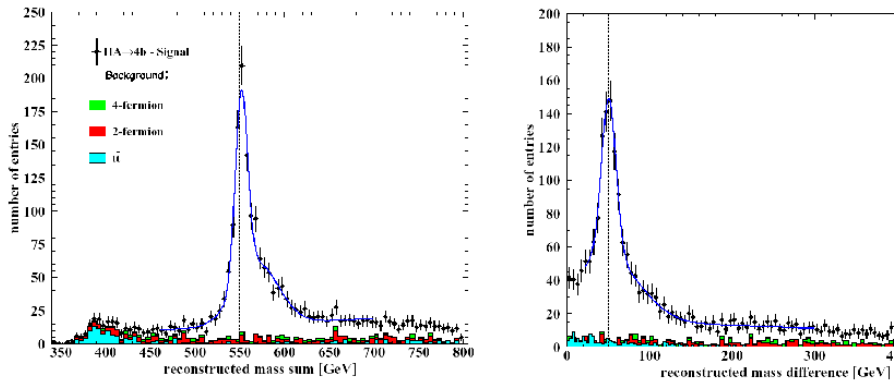,width=12cm}}
\end{picture}
\end{center}
  \caption[]{\label{fig:higgssusy} Reconstructed mass sum and difference for $HA \rightarrow 4b$ for masses of $m_H = 250$~GeV and $m_A = 300$~GeV at a center of mass energy of 800~GeV~\cite{tdr}.}
\end{figure}

At the ILC, associated $hA$ and $HA$ production allows very precise measurements of the SUSY Higgs bosons. As an example, figure~\ref{fig:higgssusy} shows the reconstructed mass sum and difference for $HA \rightarrow 4b$ for masses of $m_H = 250$~GeV and $m_A = 300$~GeV at a center of mass energy of 800~GeV. The determination of the spin and the CP quantum numbers of the Higgs bosons via their threshold behaviour, their production angles or even via decays into $\tau$ pairs will then allow to disentangle the various Higgs bosons. 
 
If there is no light Higgs boson found at the LHC, then the serious questions is: ``what makes todays precision observables behave as if there were a light Higgs in the loops?" In this case, it is especially important to use the ILC's precision to probe virtual effects. For those, the sensitivity of the ILC reaches far beyond its center of mass energy into the multi--TeV range, in many cases substantially beyond even the LHC reach. 

\section{Dark Matter \& Extra Dimensions}

\begin{figure}[htb]
\begin{center}
\begin{picture}(9,5)
\put(0,0){\epsfig{figure=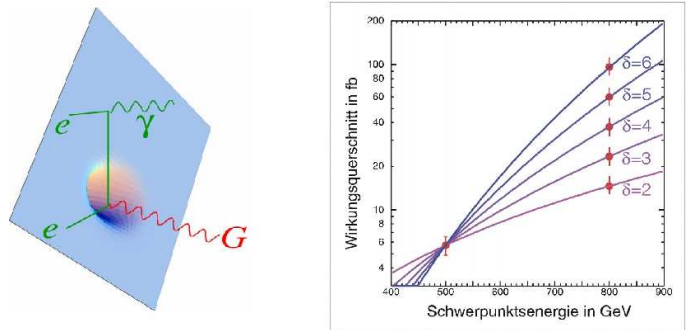,width=9cm}}
\end{picture}
\end{center}
  \caption[]{\label{fig:extra} Left: Production mechanism for $e^+e^- \rightarrow \gamma G$, the graviton escapes into extra dimensions. Right: The cross section of this process as a function of the center of mass energy for different numbers of extra dimensions. The points with error bars indicate the precision of the ILC measurements~\cite{tdr}.}
\end{figure}

The notion of additional space-time dimensions appears in many theories beyond the Standard Model of particle physics. Especially models in which only gravity can propagate in the extra dimensions, while all other particles and interactions are confined in the usual four dimensional world, are very attractive because they could explain why gravity appears to be so much weaker than the other interactions. Inother words, it provides a solution to the problem of the large hierarchy between the electroweak and the Planck scale. If such a model is realised in nature, then some of the gravitons produced at the ILC via $e^+e^- \rightarrow \gamma G$ will escape into the extra dimension,
producing events with a photon and missing energy and momentum w.r.t. to the initial beam particles. The energy dependence of the cross section for this process is closely correlated to the number of extra dimensions $\delta$. The main background for such processes is SM neutrino production with an additional hard initial state photon, which can be suppressed by one order of magnitude by chosing the right beam polarisations, allowing thus to determine the number of extra dimensions as shown in figure~\ref{fig:extra}.

\begin{figure}[htb]
\begin{center}
\begin{picture}(10,5)
\put(0,0){\epsfig{figure=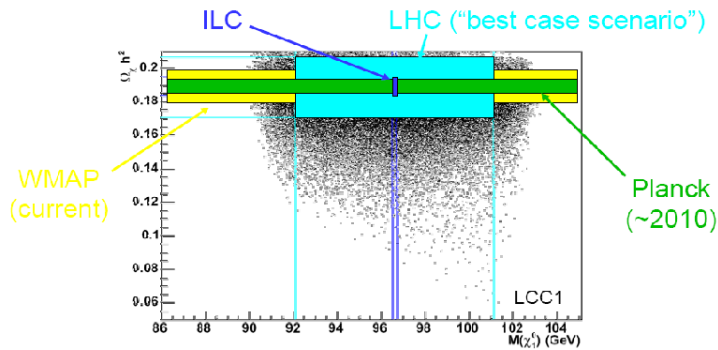,width=10cm}}
\end{picture}
\end{center}
  \caption[]{\label{fig:dm} Dark matter relic density vs WIMP mass in the LCC1 SUSY scenario. Possible paramter choices are indicated as black dots, which are compared to the sensitivity of present and future measurements from satelite and accelerator based experiments~\cite{wimp}.}
\end{figure}

Another intriguing topic for the ILC is the understanding of the dark matter which makes up 23\% of our universe according to recent cosmological observations\cite{wmap}. The Standard Model of particle physics does not contain a suitable candidate to explain this large amount. One type of candidates which is predicted in many models beyond the SM are weakly interacting massive particles (WIMPs). In Supersymmetry (SUSY) for example, the lightest supersymmetric particle (LSP) can be such a WIMP candidate, provided it is electrically neutral, not coloured and stable, i.e. if R--parity is conserved. In many SUSY scenarios the lightest neutralino plays this role, which would be discovered at the LHC. The observation and even the mass measurement alone however do not answer the question whether the LSP accounts for the dark matter in the universe. This can only be clarified when all relevant parameters of the model are determined, so that the cross section for all reactions of the WIMPs with themselves and/or other particles and thus the relic density can be calculated. This is illustrated in figure~\ref{fig:dm}, which shows the calculated relic density as a function of the mass of the lightest neutralino in the so called LCC1 scenario. The yellow band shows the current knowledge of the relic density from the WMAP experiment, which is expected to be improved significantly by the Planck satellite in the future (green band). The scattered points result from a scan over the free parameters of this scenario, which can for the same LSP mass yield a large variation in the relic density. The light blue and dark blue boxes indicate the precision of the measurements at the LHC and ILC, respectively. It is evident that only the precision of the ILC can match in an appropriate way the precision of the cosmological observations.

Another unique opportunity for dark matter searches at the ILC opens up independently from specific models. By only assuming that WIMPs annihilate into SM particles, one can use the observed relic density and crossing symmetries to calculate an expected rate for WIMP pair production via $e+e^- \rightarrow \chi \chi$.  The predicted rates\cite{birkedal} show that such events could be observed at the ILC by the detection of an additional hard initial state radiation photon, however further more detailed studies are needed.

\section{Conclusions}

New physics phenomena related to the electroweak symmetry breaking are likely to appear at the TeV scale and new particles will then be observed at the LHC. However, only high precision measurements can unveil the underlying theories at higher scales, as for example the question if the strong, electronmagnetic and weak couplings unify indeed near the GUT scale. Maybe the extrapolation of couplings and masses will be the only experimental clue to GUT scale physics.

The physics case for a (sub-)TeV electron positron collider running in parallel with the LHC is compelling. The ILC will be ideally suited to map out the profile of the Higgs boson -- or what ever takes its role - and provide a telescopic view to physics at highest energy scales. 
Furthermore, the ILC will make significant contributions to the understandig of cosmological questions like the nature of dark matter.
The detector for such a precision machine is a challenge. Conceptual detector design choices need to be made in a few years time and must be prepared now in parallel to the ongoing R\&D for the single detector components.

\section*{Acknowledgments}
I thank my colleagues Felix Sefkow and Klaus Desch for their valuable input to this talk.

\end{document}